\documentclass[aps,prr,twocolumn,groupedaddress,longbibliography,showpacs,amsmath,amssymb,amsfonts,nofootinbib]{revtex4-1}
\usepackage{graphicx}
\usepackage{dcolumn}
\usepackage{bm}
\usepackage[usenames,dvipsnames]{color}
\usepackage{multirow}
\usepackage{hyperref}
\usepackage{soul}
\usepackage{natbib}
\usepackage{amsmath, amssymb}
\usepackage{bbold}
\usepackage{physics}
\usepackage{cancel}
\usepackage{hyperref}

\definecolor{magenta}{rgb}{0.8,0.2,0.8}

\begin{document}

\title{
Kardar-Parisi-Zhang fluctuations in the synchronization dynamics of limit-cycle oscillators}

\author{Ricardo Guti\'errez }
\author{Rodolfo Cuerno}
\affiliation{Grupo Interdisciplinar de Sistemas Complejos (GISC), Departamento de Matemáticas, Universidad Carlos III de Madrid, 28911 Legan{\'e}s, Madrid, Spain}

\begin{abstract}
The time-dependent process whereby one-dimensional systems of self-sustained oscillators synchronize is shown to display scale invariance in space and time, akin to that found in the dynamics of equilibrium critical phenomena. Remarkably, the process is largely independent of system details, sharing with a class of nonequilibrium surface kinetic roughening the universal scaling behavior of the Kardar-Parisi-Zhang equation with columnar noise, and featuring phase fluctuations that follow a Tracy-Widom probability distribution. This is revealed by a numerical exploration of rings of Stuart-Landau oscillators (the universal representation of an oscillating system close to a Hopf bifurcation) and rings of van der Pol oscillators, both paradigmatically supporting self-sustained oscillations. The critical behavior is very well defined for limit-cycle oscillations near bifurcation, and still dominates comparatively far from it. In particular, the Tracy-Widom fluctuation distribution seems to be an extremely robust feature of the synchronization process. The nonequilibrium criticality here described appears to transcend the details of the coupled dynamical systems that synchronize, 
making plausible its experimental observation.

\end{abstract}


\maketitle

\section{Introduction}
\label{secintro}

A great variety of 
systems comprising several dynamical units exhibit, and sometimes crucially 
require, synchronous dynamics \cite{pikovsky}, 
including ensembles of neurons and heart cells, fireflights, flocking birds and humans, coupled pendula, and quantum oscillators. They may be 
described as few- or many-body systems of oscillators, with regular or chaotic dynamics, subjected to external forcing or mutual coupling, with all-to-all, periodic, or irregular interaction patterns.  All such cases present 
synchronization and have been intensively researched 
\cite{kuramoto_book,boccaletti,osipov,arenas}.

Key properties of the time-dependent process through which spatially-coupled oscillators become synchronized 
are also found in the surface kinetic roughening behavior occurring in nonequilibrium growth processes like the formation of e.g.\! thin solid films, bacterial colonies, or coffee rings \cite{barabasi,halpinhealy,krug97}. Indeed, such features are lately proving to be of a broad relevance to many non-interfacial systems, from the evolution of Lyapunov vectors in space-time chaos \cite{Pikovsky1994,Pikovsky1998,Pazo2010,Fukai2021,Odavic2021} to the one-dimensional (1D) dynamics of fermions and bosons \cite{Fujimoto2020,Fujimoto2021,Fujimoto2022}, or the kinetics of chemical reactions \cite{Mondal2022}. All these systems display universal, scale-invariant behavior as in the dynamics of equilibrium critical systems \cite{goldenfeld,tauber14}, but without requiring parameter fine tuning, 
providing instances of generic scale invariance (GSI) \cite{grinstein91,grinstein95,belitz05}. 
The Kardar-Parisi-Zhang (KPZ) equation \cite{kardar,kriecherbauer10,takeuchi} is paramount in this context, as its universality class describes the critical behavior found in many strongly correlated systems, from surface growth processes \cite{barabasi,halpinhealy,krug97} to novel non-interfacial contexts like active \cite{chen16} or quantum \cite{wei22,fontaine,Mu2024} matter.
The ubiquity of KPZ universality seems connected to the fact that the statistics of fluctuations in those systems are governed \cite{kriecherbauer10,takeuchi} by the (non-Gaussian) Tracy-Widom (TW) family of probability density functions (PDF) for random matrix eigenvalues \cite{tracy09}, which have lately been recreating the generality of Gaussian statistics, but for a host of low-dimensional systems with strong correlations \cite{Fortin2015,makey20}. 

In the synchronization context, a formal connection with surface kinetic roughening has been pointed out repeatedly in the literature \cite{sakaguchi,kuramoto_book,grinstein,chate1996,manneville1996,pikovsky,Lepri2022}.
For lattices of {\em idealized phase oscillators}, we have clarified this connection \cite{gutierrez}, showing that the synchronization process displays GSI whose critical behavior falls into specific universality classes of kinetic roughening. 
Namely, for odd-symmetric coupling functions, as for the celebrated Kuramoto model \cite{acebron}, the nonlinear effects which are the hallmark of the KPZ class play no role \cite{gutierrez}, and the scaling behavior is that of the linear Edwards-Wilkinson (EW) equation with columnar quenched disorder, also known as the Larkin model \cite{purrello}, with Gaussian fluctuations. Otherwise, for generic coupling functions the KPZ nonlinearity is at play, inducing the scale-invariant behavior of the columnar KPZ equation \cite{szendro} and TW fluctuations \cite{gutierrez}. These results depend on two strong assumptions, however: I) the dynamical units are phase oscillators (i.e.~idealized self-sustained oscillators), and II) the pairwise coupling is purely sinusoidal.

\begin{figure}[t!]
\includegraphics[scale=0.45]{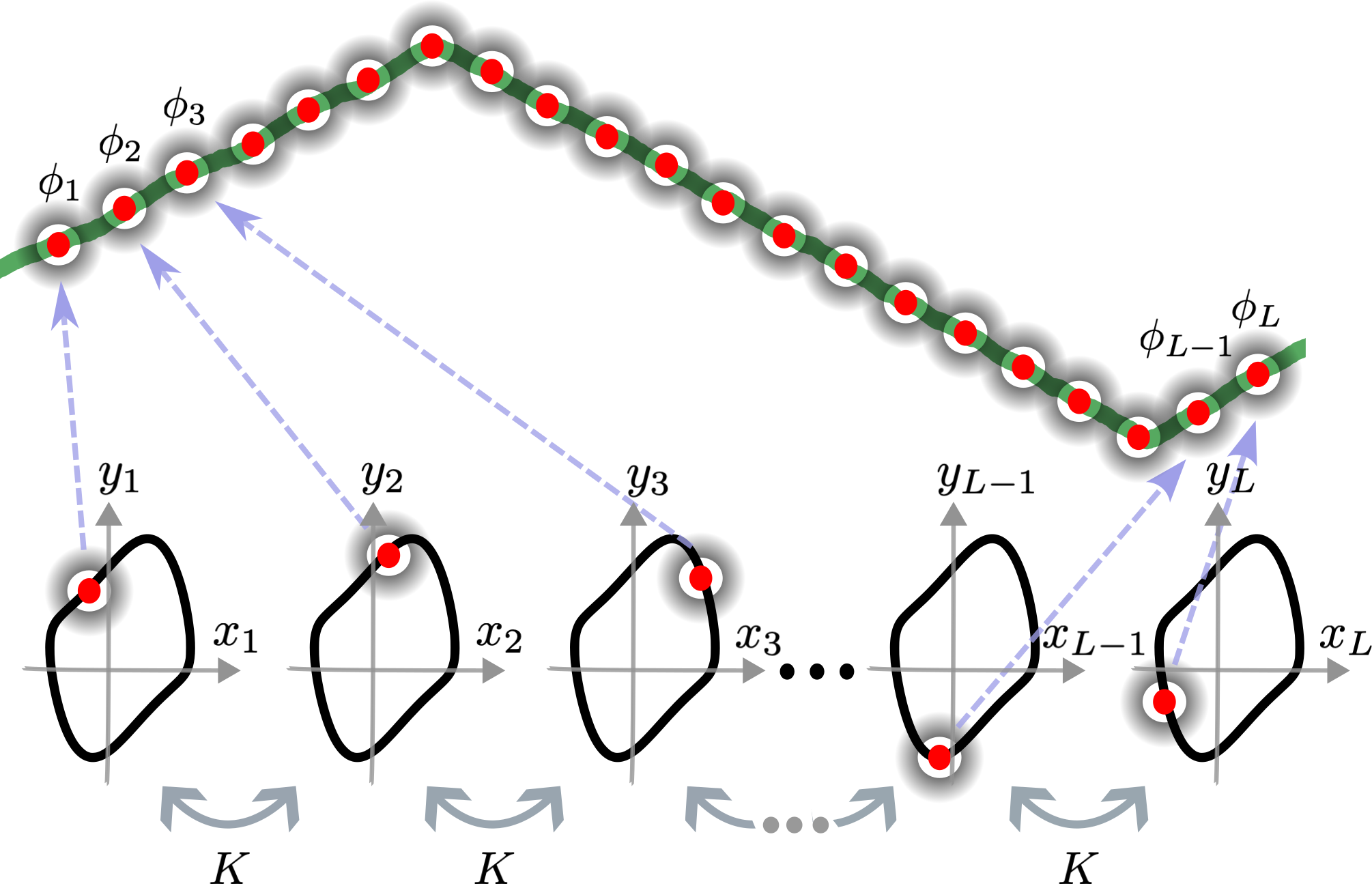}
\vspace{0cm}
\caption{
Phases $\phi_j(t)$ extracted from the states of oscillators form an ``interface'' above an oscillator ``substrate'', where the subindex $j$ identifies an oscillator by its position along the lattice. The limit cycle shown is for a single vdP oscillator with $\mu = 1$, Eq.\ (7), $K=0$, and the green interface is a snapshot of simulations of the same equation for $K\neq 0$.} 
\label{fig1}
\end{figure}

In this article, we elucidate the universal features of the synchronization process 
in lattices of full-fledged, self-sustained oscillators of the Stuart-Landau (SL) and van der Pol (vdP) types. While the SL model is a canonical representation of an oscillating system close to its Hopf bifurcation \cite{kuramoto_book}, the vdP model is a paradigmatic system of experimental relevance \cite{pikovsky,strogatzbook}. We thus bypass the limitations of Ref.~\cite{gutierrez} by studying generic limit-cycle oscillators, with coupling functions that are not necessarily sinusoidal at the phase-reduced level \cite{pietras}. To do so, we adapt the observables and analyses employed to assess universal behavior in kinetic roughening to the study of the phase dynamics \cite{pikovsky} of coupled SL or vdP oscillators, the phases themselves playing a role analogous to that of local heights in interfacial systems \cite{barabasi,halpinhealy,krug97}. Intuitively, a ``phase interface'' grows above a ``substrate'' of oscillators, an analogy fruitfully explored in various recent works dealing with phase oscillators \cite{lauter,moroney,gutierrez}; see Fig.~\ref{fig1} for an illustration. Our results show that GSI is robust for the synchronization process, generically displaying the dynamic scaling Ansatz and exponent values of the columnar KPZ equation, and TW-distributed fluctuations, as for idealized phase oscillators with non-odd coupling \cite{gutierrez}. This makes plausible the experimental observation of such nonequilibrium criticality in synchronous dynamics.

The paper is organized as follows: In Sec.\ \ref{secSL} we first analyze the synchronization process of 1D lattices where each site is a SL oscillator, ``an ideal nonlinear oscillator model'' \cite{kuramoto_book}. Then in Sec.\ \ref{secvdP} we address physically-motivated limit-cycle oscillators, here vdP oscillators, which are shown to display essentially the same scaling behavior in their synchronization process as SL (and phase \cite{gutierrez}) oscillators. 
Finally, in Sec.\ \ref{secConclusions} we provide our conclusions and some ideas for the experimental observation of such nonequilibrium criticality properties in synchronous dynamics. Additionally, we present a detailed analytical discussion of the phase-reduced dynamics for SL and vdP oscillators in Appendices \ref{AppA} and \ref{AppB}, respectively. Appendix \ref{AppC} provides further numerical results on real-space two-point correlations and phase ``interfaces" for SL and vdP oscillator rings, while Appendix \ref{AppD} details our numerical definition of the dynamical phase for the latter.

\section{Stuart-Landau oscillators} 
\label{secSL}

A SL oscillator is the normal form of a nonlinear dynamical system undergoing a supercritical Hopf bifurcation, where a fixed point loses stability and a small-amplitude limit cycle branches from it \cite{nakao,introsl}. We consider a 1D chain of $L$ such oscillators with periodic boundary conditions (PBC, i.e.\! a ring), namely,
\begin{align}
\dot{z}_j(t) &= a_j z_j(t) - b_j |z_j(t)|^2 z_j(t)\nonumber \\
&+ K c_j [z_{j+1}(t) + z_{j-1}(t) - 2 z_j(t)],
\label{SL}
\end{align}
where a dot denotes time derivative and $j=1,2,\ldots, L$. The $j$-th oscillator is described by the complex-valued function $z_j(t)$, while $a_j = \mu + i\, \omega_j$, $b_j = 1$, $c_j = 1+i \in \mathbb{C}$, and the coupling strength $K \in \mathbb{R}$, are constants. Here, $\mu$ measures the distance to the ($\mu_c=0$) bifurcation point, and the intrinsic frequency $\omega_j$ is oscillator-dependent. The first line on the right-hand side of Eq.~(\ref{SL}) contains the standard SL-oscillator dynamics \cite{introsl}, and the second line adds a (diffusive) nearest-neighbor coupling, where $z_{0}(t) \equiv z_{L}(t)$ and $z_{L+1}(t) \equiv z_1(t)$. Numerical evidence suggests that the GSI remains unchanged, for both SL and vdP oscillators, if open boundary conditions are used instead of PBC (not shown).

\begin{figure*}[t!]
\includegraphics[scale=0.39]{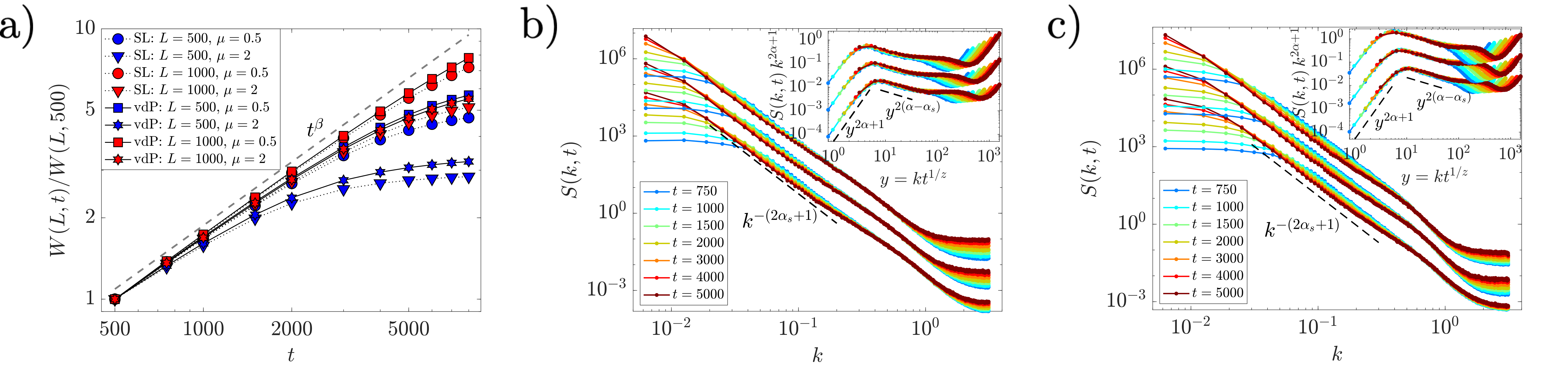}
\vspace{-0.2cm}
\caption{
(a) Roughness normalized with respect to its value at $t=500$, $W(L,t)/W(L,500)$, as a function of time for different sizes $L$ in rings of SL oscillators and rings of vdP oscillators. Two choices of the parameter $\mu$ are displayed (see legend). The dashed line represents a power-law growth $t^{\beta}$. Panels (b) and (c): Structure factor $S(k,t)$ at times as in the legend, for rings of $L=1000$ SL oscillators [(b), main panel] and of vdP oscillators [(c), main panel]. In both panels the insets show the rescaling of the $S(k,t)$ data of the main panel following Eq.~(\ref{Skscal}). Results are for $\mu = 0.5, 1$, and $2.0$, the last two values shown after an upward vertical displacement, so that larger values of $\mu$ appear above, for visibility purposes. In all three panels, the exponent values used are $\alpha = 1.07$, $z = 1.37$, and $\alpha_s = 1.40$ ($\beta = \alpha/z \approx 0.78$). Results based on $1000$ realizations.}
\label{fig2}
\end{figure*}

In the uncoupled case ($K = 0$), each oscillator has a circular stable limit cycle on the complex plane given by $z_j(t) = R_j(t) \exp{i \theta_j(t)}$, with radius $R_j(t) = \sqrt{\mu}$ and geometric phase $\theta_j(t) = \omega_j t$ \cite{introsl}, the latter being a valid dynamical phase variable \cite{pikovsky}. 
Outside the limit cycle (for $R_j(t) \neq \sqrt{\mu}$), an asymptotic phase \cite{kuramoto_book} can be explicitly defined as $\phi(z_j(t)) = \theta_j(t)$ (i.e.\! the dynamic phase is the geometric phase 
also there), which provides a set of isochrones covering phase space, such that on each one of them $\dot{\phi}(z_j(t)) = \omega_j$ \cite{nakao}. One can thus easily extract the instantaneous phases $\{\phi_j(t)\}_{j=1}^L$ from the states $\{z_j(t)\}_{j=1}^L$ (on or off the limit cycle) of the oscillators whose dynamics are governed by Eq.~(\ref{SL}).

In our simulations, the intrinsic frequencies $\{\omega_j\}_{j=1}^L$ are chosen at random from a uniform distribution on an interval of width $\epsilon$ centered at 
$1$, i.e.\! $\omega_j \sim U([1-\epsilon/2, 1+\epsilon/2])$. The ensuing heterogeneity is akin to that found in phase-oscillator models like the Kuramoto model \cite{acebron}. In the kinetic roughening literature  \cite{halpinhealy} this is known as {\it columnar noise}, i.e.\! quenched disorder that only depends on the substrate position (oscillator), not on the height (phase) value. In practice, we take $\epsilon = 0.1$, while we inspect various values of $\mu$, which also determines the coupling strength, $K=\mu$. The equation set (\ref{SL}) is integrated by a $4$-th order Runge-Kutta method with step size $0.01$ ---to be employed for our vdP oscillators too--- from the initial condition $z_j(0) = \sqrt{\mu}$ for all $j$.

The deviations of the local phases around their mean value are measured by the global width or roughness,
\begin{equation}
W(L,t) \equiv \langle \overline{[\phi_j(t)-\overline{\phi_j(t)}]^2} \rangle^{1/2},
\label{W}
\end{equation}
with the overbar denoting spatial average in a substrate of size $L$ and the brackets denoting average over different noise realizations, as is standard for kinetically rough interfaces \cite{barabasi}.
Here $W$ saturates if the phase differences remain bounded in time, which only happens when all oscillators reach a common effective frequency. Thus, saturation manifests synchronization (i.e.\! frequency entrainment) \cite{gutierrez}, only occurring for coupling strengths $K$ above the threshold value $K_c$. In surface kinetic roughening, the heights (here, the phases) are statistically correlated for distances below a correlation length $\xi(t)$ increasing with time as $\xi(t) \sim t^{1/z}$, where $z$ is the so-called dynamic exponent. This behavior holds until $\xi(t)$ reaches a value comparable to $L\gg 1$, implying saturation of the width at a steady-state, size-dependent value $W(L,t\gg L^z) \sim L^\alpha$, where the roughness exponent $\alpha$ characterizes the fractality of the phase profile $\{\phi_j(t)\}_{j=1}^L$. Prior to saturation, the roughness increases in time as $W(t)\sim t^{\beta}$, where $\beta=\alpha/z$ \cite{barabasi,halpinhealy,krug97}.

In Fig.~\ref{fig2}(a) we show the roughness $W(L,t)$ for systems of $L = 500$ and $1000$ SL oscillators. For each $L$, we consider two different values $\mu = 0.5$ and $2$, both of which yield $K = \mu > K_c$, so that synchronization is achieved, which as expected occurs later for larger systems. Before saturation, a growth regime is observed in which $W\sim t^{\beta}$ closely, for $\beta \approx 0.78$. This corresponds to the choice of exponent values $\alpha = 1.07$ and $z = 1.37$,  justified below.


Beyond the roughness, kinetic roughening also manifests itself in the space-time behavior of correlation functions \cite{barabasi,halpinhealy,krug97}. A very useful one \cite{siegert96,lopezphysa} is the power spectral density of phase fluctuations,
\begin{equation}
S(k,t) \equiv \langle \hat{\phi}(k, t)\hat{\phi}(-k, t)\rangle = \langle |\hat{\phi}(k, t)|^2\rangle,
\label{Sk}
\end{equation}
where $\hat{\phi}(k,t)$ is the (1D) space Fourier transform of $\phi_j(t)$ and $k$ is the wavenumber. As a generalization of equilibrium critical behavior \cite{tauber14}, in kinetic roughening this correlation function satisfies the scaling form \cite{ramasco}
\begin{equation}
S(k,t) = k^{-(2  \alpha +1)} s(k t^{1/z}) ,
\label{Skscal}
\end{equation}
where the scaling function $s(y) \propto y^{2(\alpha - \alpha_s)}$ for $y\gg 1$ and $s(y) \propto y^{2\alpha +1}$ for $y\ll 1$. Here,
an additional (spectral roughness) exponent $\alpha_s$ appears, which characterizes length scales smaller than $\xi(t)$, as Eq.\ \eqref{Skscal} implies that $S(k,t)\sim 1/k^{2\alpha_s+1}$ for $k\gg 1/t^{1/z}$. In the simplest cases, as for the KPZ equation with time-dependent noise, $\alpha_s=\alpha$ and Eq.\ \eqref{Skscal} reduces to the celebrated Family-Viscsek dynamic scaling Ansatz \cite{Family1985,barabasi,halpinhealy,krug97}. However, in our
case $\alpha_s \neq \alpha$ are independent exponents such that $\alpha_s, \alpha>1$, a behavior known as faceted anomalous scaling  \cite{ramasco}. Indeed, Fig.~\ref{FigA2} in Appendix \ref{AppC} illustrates the formation of triangular facets in the phase ``interface'' along the evolution corresponding to Fig.\ \ref{fig2}(b). The latter figure shows the structure factor in a ring of $L=1000$ SL oscillators, for $\mu = 0.5, 1$, and $2$. An upward vertical displacement, larger for higher $\mu$, has been applied to the last two values to ease visualization. The main panel shows $S(k,t)$ as a function of $k$ for several times, while the inset shows the same data in rescaled form according to Eq.~(\ref{Skscal}), with $\alpha=1.07$, $z=1.37$, and $\alpha_s = 1.40\neq \alpha$ \cite{err_exp}. The data collapse is very good except for wavenumbers close to the discretization cutoff. A peculiar implication of these exponent values is that two-point correlations (related with $S(k,t)$ via Fourier transforms \cite{krug97}) scale with a different roughness exponent in real space \cite{schroeder93,dassarma,krug97,lopez,ramasco,cuerno04}. E.g., by defining $G(r,t)=\langle \overline{(\phi_{j+r}(t)-\phi_{j}(t))^2}\rangle$, under faceted anomalous scaling \cite{ramasco} $G(r)\sim r^{2\alpha_\text{loc}}$, where $\alpha_\text{loc}=1\neq\alpha\neq\alpha_s$. For the simulations in Figs.\ \ref{fig2}(a)-(b), $\alpha_\text{loc} = 0.97$ (see Appendix \ref{AppC}), the difference with the expected value possibly being a finite-size effect.

\begin{figure}[t!]
\includegraphics[scale=0.36]{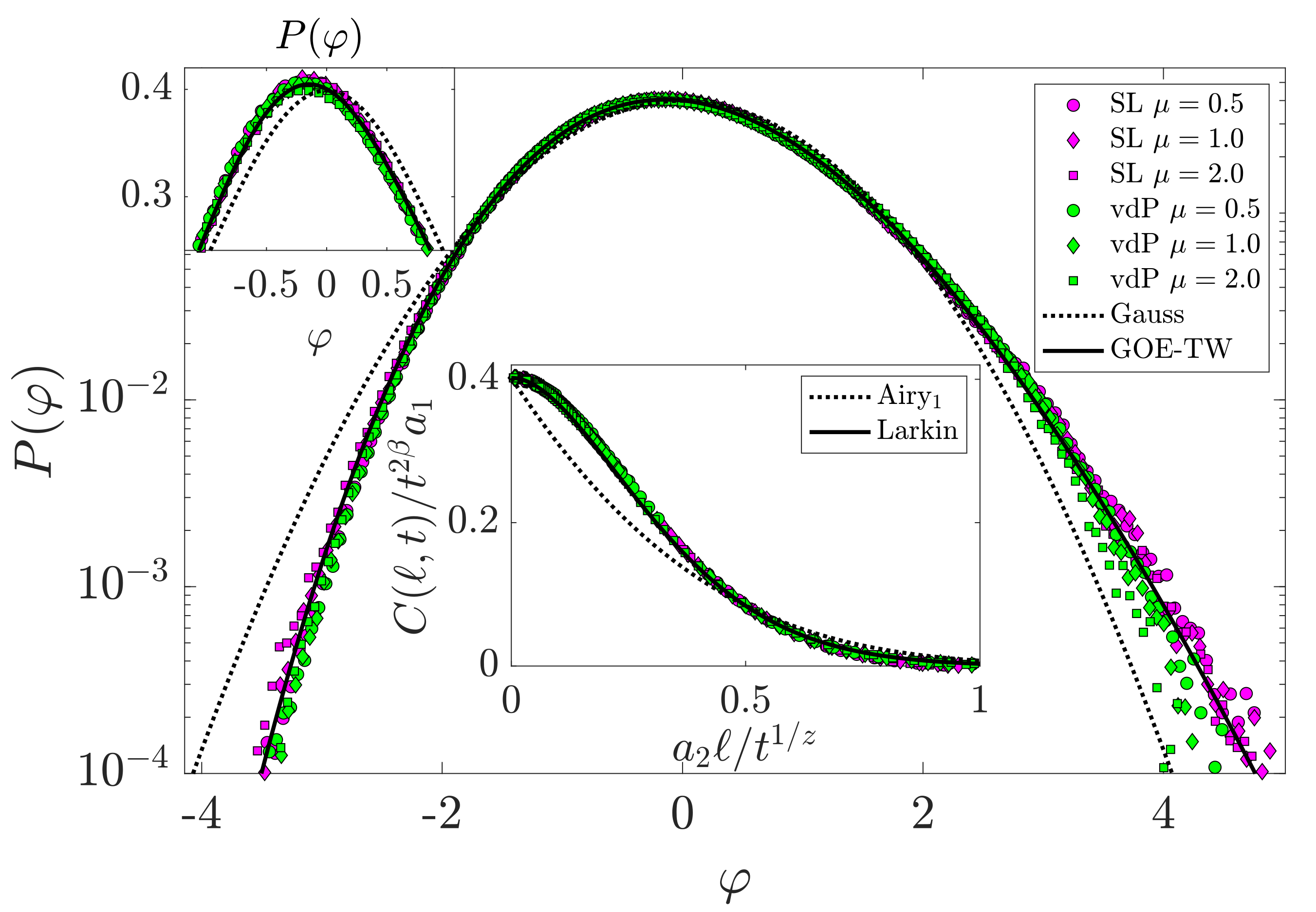}
\caption{
(Main panel) Histogram of phase fluctuations, Eq.~(\ref{fluct}), in rings of $L=1000$ SL oscillators and vdP oscillators for $\mu = 0.5, 1$, and $2$, normalized to zero mean and unit variance.  (Top-left inset) Same data in linear scale, centered around the peak of the distribution.  The dotted (solid) line represents a Gaussian (GOE-TW) PDF. Data for the histograms are based on $10000$ realizations, for $t= 300, 500, 700, 1000, 1500$, and $2000$ ($t_0 = 100$, and $\Delta t = 200, 400, \ldots, 1900$), in the growth regime. (Bottom-center inset) Rescaled phase covariance, Eq.~(\ref{eq:cov}), for rings of SL and vdP oscillators with parameter values as in (a), for $t = 500, 550, 600,\ldots, 800$. The dotted (solid) line represents the covariance of the Airy$_1$ process (Larkin model). The exponent values are $\alpha = 1.07$ and $z = 1.37$ ($\beta = \alpha/z \approx 0.78$).}
\label{fig3}
\end{figure}

As recently highlighted for e.g.\ KPZ-related systems \cite{kriecherbauer10,takeuchi}, the statistics of phase fluctuations $\phi_j(t)$ around their mean is another trait of a kinetic roughening universality class. We thus study the PDF of
\begin{equation}
\varphi_j\equiv \frac{\delta \phi_j(t_0+\Delta t) - \delta \phi_j(t_0)}{(\Delta t)^{\beta}} ,
\label{fluct}
\end{equation}
where $j =1,2,\ldots, L$ and $\delta \phi_j(t) = \phi_j(t) - \overline{\phi_j(t)}$; $t_0$ is a reference time past initial transients, with $t_0 +\Delta t$ an intermediate time within the growth regime. Division by $(\Delta t)^{\beta}$ removes the systematic increase of the fluctuations in time so that, remarkably, the PDF of $\varphi_j$ reaches a universal, time-independent form \cite{kriecherbauer10,takeuchi}. Examples in the kinetic roughening context are the Gaussian distribution for the EW equation \cite{barabasi,krug97} and a TW PDF (depending, e.g., on boundary conditions) for the KPZ equation \cite{kriecherbauer10,takeuchi}. In lattices of phase oscillators, the former is observed for Kuramoto coupling and the latter for generic coupling functions \cite{gutierrez}. For rings of SL oscillators, Fig.\ \ref{fig3} shows the histograms of fluctuations $\varphi_j$ for $\mu = 0.5,1,2$, focusing on the tails (main panel) and central part (top-left inset). 
Data follow quite closely the Gaussian Orthogonal Ensemble (GOE)-TW PDF, as for the 1D KPZ equation with time-dependent noise and PBC \cite{kriecherbauer10,takeuchi}.

Finally, two-point fluctuation statistics are also universal and can be assessed \cite{takeuchi} through the phase covariances
\begin{equation}
C(r,t) \equiv \langle \overline{\phi_{j}(t) \phi_{j+r}(t)} \rangle - \langle \bar{\phi}(t) \rangle^2.
    \label{eq:cov}
\end{equation}
They are shown in the bottom-center inset of Fig.~\ref{fig3} for the same three values of $\mu$. In the growth regime \cite{kriecherbauer10,takeuchi}, $C(\ell,t) = a_1 t^{2\beta} \mathcal{C}(a_2 \ell/t^{1/z})$, where $\mathcal{C}$ is a scaling function, $\ell \equiv |r|$, and $a_1$ and $a_2$ are constants. To fix them \cite{barreales2020}, we have chosen the covariance of the Airy$_1$ process, which occurs in the 1D KPZ class with PBC \cite{takeuchi}, as the functional form for $\mathcal{C}(x)$. Instead, the results follow quite closely the covariance of the Larkin model \cite{purrello,gutierrez}. Figure \ref{fig3} for the SL rings thus agrees with the fluctuation PDF and phase covariance that we have recently found \cite{gutierrez} for phase oscillators with a generic (non-odd) sinusoidal coupling function, as analytically implied by a phase reduction study, see Appendix \ref{AppA}. Yet the occurrence of such PDF and covariance is far from trivial, as the coupling between the oscillators is relatively strong, the differences in their intrinsic frequencies are sizable, and the distance to the bifurcation point considerable, while phase reduction works best for almost identical weakly-coupled oscillators close to the bifurcation \cite{pietras}. For a still more radical departure from idealized regimes, we next study a physically-motivated oscillator which behaves very differently far from its bifurcation threshold.

\section{van der Pol oscillators} 
\label{secvdP}

The vdP oscillator has played a central role in the development of nonlinear science \cite{pikovsky,strogatzbook,misbah}.
We analyze a ring of $L$ coupled oscillators (generalizing the $L=2$ case \cite{Rand80}), whose evolution reads
\begin{align}
\ddot{x}_j &= (\mu - x_j^2)\, \dot{x}_j - \omega_j^2\, x_j
\\ &+ K[(x_{j+1} + x_{j-1} - 2 x_j)+(\dot{x}_{j+1} + \dot{x}_{j-1} - 2 \dot{x}_j)],\nonumber
\label{eq:vdP}
\end{align}
where $j = 1,2,\ldots, L$. The state of the $j$-th oscillator is given by the pair of real numbers $(x_j(t), \dot{x}_j(t))$, with $x_{0}(t) \equiv x_{L}(t)$ and $x_{L+1}(t) \equiv x_1(t)$. The first two terms on the right-hand side are those of the vdP oscillator, the remaining one being a diffusive coupling term with strength $K$. In the absence of coupling ($K=0$), the $j$-th dynamical unit oscillates with amplitude proportional to $\sqrt{\mu}$ at an intrinsic frequency $\omega_j$ for $\mu>0$. 
The critical value $\mu_c = 0$ is the bifurcation point of a supercritical Hopf bifurcation \cite{strogatzbook}. We explore different values of $K=\mu > K_c > 0$, so that synchronization occurs in all cases. The intrinsic frequencies $\{\omega_j\}_{j=1}^L$ are again random, taken from a uniform distribution, $\omega_j \sim U([0.95, 1.05])$.

For sufficiently small $\mu$ one might expect a similar behavior to that found for our SL oscillators, Eq.\ \eqref{SL}.
As $\mu$ is increased, however, the vdP oscillator behaves more as a relaxation oscillator, moving across different regions of the limit cycle (which differs from a circle, see Fig.~\ref{fig1} for $\mu=1$) at different speeds \cite{strogatzbook}.
Thus, the geometric $\arctan(\dot{x}_j/x_j)$ phase 
no longer behaves as a valid dynamical phase variable \cite{pikovsky} 
and the phase $\phi_j(t)$ is instead defined by a Poincar\'e section \cite{pikovsky}, see Appendix \ref{AppD} for details.

In Fig.~\ref{fig2}(a) we show the roughness $W(L,t)$ for rings of vdP oscillators in which $\mu = 0.5$ and $2$, ranging from close to $\mu_c$, well into relaxation-oscillation values of $\mu$.
Figure \ref{fig2}(c) displays the structure factor $S(k,t)$ across time, vertically shifted as in Fig.\ \ref{fig2}(b).
These data are collapsed in the inset, according to Eq.\ \eqref{Skscal}, using the same exponent values as for the SL oscillators. Again, collapse is very good except for $k$ values close to the discretization cutoff.


In Fig.~\ref{fig3} (main panel and top-left inset), the distribution of fluctuations $\varphi_j$ in the system of vdP oscillators is shown to follow quite closely the GOE-TW PDF, as for SL oscillators. This might be expected close to bifurcation ($\mu \ll 1$), as phase reduction should be accurate there, and the coupling function of the phase oscillators is not an odd function (which would imply Gaussian fluctuations \cite{gutierrez}), see Appendix \ref{AppB}. 
Actually, we find that the GOE-TW fluctuation PDF is a robust feature of synchronization that holds even for large $\mu$. 
The same can be said about the phase covariance $C(r,t)$, shown in Fig.~\ref{fig3} (bottom-center inset), which again closely follows the covariance of the Larkin model \cite{gutierrez}.

\section{Conclusions} 
\label{secConclusions}

Overall, for the two paradigmatic 1D systems of self-sustained oscillators that we study, the gradual emergence of synchronization displays GSI compatible with the faceted scaling of the columnar KPZ equation, a TW PDF, and a Larkin covariance. The fact that the same critical behavior was recently observed for idealized phase oscillators with harmonic couplings \cite{gutierrez} is not obvious, since first-order phase-reduced models may fail to capture the dynamical behavior of limit-cycle oscillators (see Ref.~\cite{Mau} for a recent example concerning chimera states). The columnar KPZ equation is prone to nonuniversal corrections \cite{krug97,szendro,nattermann,krughh}, expected here from changes in the $\{\omega_j\}_{j=1}^L$ distribution, which may justify the small differences in our critical exponent values with respect to those in Ref.~\cite{szendro}. Still, the behavior we observe is robust and occurs even far from bifurcation. Given the experimental relevance of e.g.\ vdP oscillators \cite{pikovsky,Itoh2001,Choubey2010,misbah}, our results make the observation of such nonequilibrium criticality, e.g.\! for coupled nonlinear electronic circuits or oscillating chemically-reacting media, an alluring possibility.



\begin{acknowledgments}
\noindent {\it Acknowledgements--} This work has been partially supported by Ministerio de Ciencia e Innovaci\'on (Spain), by Agencia Estatal de Investigaci\'on (AEI, Spain, 10.13039/501100011033), and by European Regional Development Fund (ERDF, A way of making Europe) through Grants No.\ PID2021-123969NB-I00 and No.\ PID2021-128970OA-I00, and by Comunidad de Madrid (Spain) under the Multiannual Agreement with UC3M in the line of Excellence of University Professors (EPUC3M23), in the context of the V Plan Regional de Investigaci\'on Cient\'{\i}fica e Innovaci\'on Tecnol\'ogica (PRICIT).
\end{acknowledgments}

\appendix
\section{Phase reduction of a chain of Stuart-Landau oscillators}
\label{AppA}

The dynamics of a ring 
comprising $L$ diffusively-coupled SL oscillators is given by
\begin{align}
\dot{z}_j(t) &= a_j z_j(t) - b_j |z_j(t)|^2 z_j(t) \\ 
&+ K c_j [z_{j+1}(t) + z_{j-1}(t) - 2 z_j(t)],\nonumber
\label{SLapp}
\end{align}
for $j = 1,2,\ldots, L$, where $z_{L+1} \equiv z_1$ and $z_{0} \equiv z_L$. In the following we focus on a reference oscillator, for which we drop the subindex $j$, and replace $z_{j+1}$ by $z_r$ (right neighbor) and $z_{j-1}$ by $z_\ell$ (left neighbor). Except for the (real) coupling strength $K$, all parameters are complex, $a = a_R + i a_I$, $b= b_R + i b_I$ and $c = c_R + i c_I$, thus
\begin{align}
\dot{z}(t) &= (a_R + i a_I) z(t) - (b_R + i b_I) |z(t)|^2 z(t)  \\
&+ K (c_R + i c_I) [z_r(t) + z_\ell(t) - 2 z(t)]. \nonumber
\label{SLref}
\end{align}
In the absence of coupling, $K=0$, the system has a limit cycle solution of radius $R_c = \sqrt{a_R/b_R}$, with intrinsic frequency $\displaystyle \omega = a_I - b_I a_R/b_R$, as it is easy to check by trying a solution of the form $z(t) = R(t) e^{i\phi(t)}$. For the parameter choices considered in Sec.\ \ref{secSL}, $a = \mu + i \omega$, $b = 1$ and $c = 1+i$, which yields a stable limit cycle of radius $\sqrt{\mu}$ with oscillation frequency $\omega = a_I$ (which changes from one oscillator to another).

It has been shown \cite{pietras,kori} that the equation above can be reduced to a phase model for coupling $K = \mu \kappa$, of order $\mathcal{O}(\mu)$, when the trajectory of $z(t)$ deviates slightly from the unperturbed limit cycle. In such a case, the solution can thus be approximated by $z(t) \approx R(t) e^{i\phi(t)}$, where the phase $\phi(t)$ evolves according to
\begin{equation}
\dot{\phi} = \omega + \mu \kappa \mathbf{Z}(z^0) \cdot h(z^0, z^0_\ell, z^0_r) .
\label{eq:faseprev}
\end{equation}
Here, $\mathbf{Z}(z_0)$ is the so-called phase sensitivity function and $h(z^0, z^0_\ell, z^0_r)$ is the coupling function, both of which are evaluated on the limit cycle, and dot stands for the complex inner product \cite{pietras,kori}. These two functions can be written in terms of the phases \cite{kuramoto_book,kori} as
\begin{equation}
\mathbf{Z}(\phi) = \frac{-b_I/b_R + i}{R_c} e^{i\phi},
\end{equation}
and
\begin{equation}
h(\phi, \phi_\ell, \phi_r) = \mu \kappa c R_c [e^{i\phi_r} + e^{i\phi_\ell} -2 e^{i\phi}].
\end{equation}
By introducing these expressions into Eq.\ \eqref{eq:faseprev} and averaging over a full period, we are finally left with a closed equation for the phases, namely,
\begin{equation}
\dot{\phi} = \omega + \mu \kappa \left[\Gamma(\phi_r - \phi) + \Gamma(\phi_\ell - \phi)\right].
\label{eq:faseSL}
\end{equation}
The function $\Gamma(\psi)$ that couples phases to one another in Eq.\ \eqref{eq:faseSL} can be expanded in a Fourier series, $\Gamma(\psi) = A_0 + \sum_{k=1}^\infty (A_k \cos k \psi + B_k \sin k \psi)$, which yields
\begin{align}
A_0 &= -A_1 = \frac{c_R b_I}{b_R} - c_I,\\
B_1 &= c_R + \frac{c_I b_I}{b_R},
\end{align}
while all other coefficients vanish at this order of approximation. Again, for the parameters employed in Sec.\ \ref{secSL}, we obtain $A_0 = -A_1 = -1$, and $B_1 = 1$, thus the coupling function is
\begin{align}
\Gamma(\psi) = -1 + \cos\psi + \sin \psi.
\label{couplingPRSL}
\end{align}
As the coupling function is certainly not odd [$\Gamma(\psi) + \Gamma(-\psi) \neq 0$], the phase-reduced model should yield the KPZ equation with columnar noise in the continuum approximation according to the arguments given in Ref.~\cite{gutierrez}. This is compatible with the GSI behavior observed in the numerical results of Sec.\ \ref{secSL}. In fact, the phase-reduced dynamics with coupling given by Eq.~(\ref{couplingPRSL}) corresponds to one of the parameter choices of the sinusoidal (Kuramoto-Sakaguchi) coupling thoroughly investigated in Ref.~\cite{gutierrez}, namely $\delta = \pi/4$.

\section{Phase reduction of a chain of van der Pol oscillators}
\label{AppB}

The dynamics of the ring 
of $L$ diffusively-coupled vdP oscillators considered in Sec.\ \ref{secvdP} is given by
\begin{widetext}
\begin{equation}
\ddot{x}_j(t) = (\mu - x_j^2(t)) \dot{x}_j(t) - \omega_j^2 x_j(t) + K[(x_{j+1}(t) + x_{j-1}(t) - 2 x_j(t)) +(\dot{x}_{j+1}(t) + \dot{x}_{j-1}(t) - 2 \dot{x}_j(t))],
\label{vdPSM}
\end{equation}
\end{widetext}
for $j = 1,2,\ldots, L$, where $x_{L+1} \equiv x_1$ and $x_{0} \equiv x_L$. We again focus our discussion on a single oscillator, whose state is given by $x$ and $v \equiv \dot{x}$ (we omit the subscript for simplicity), while that of its left (right) neighbor is given by $x_\ell$ and $v_\ell \equiv \dot{x}_\ell$ ($x_r$ and $v_r \equiv \dot{x}_r$). The dynamics of the reference oscillator can be expressed as a system of two ordinary differential equations in the standard manner,
\begin{align}
\begin{pmatrix}\dot{x}\\ \dot{v}\end{pmatrix} &= \begin{pmatrix} 0 & 1\\ -\omega^2 & \mu \end{pmatrix} \begin{pmatrix}x\\ v\end{pmatrix} + \begin{pmatrix}0\\ -x^2 v\end{pmatrix} +\nonumber\\ 
&K\left[ \begin{pmatrix} 0 & 0\\ -2 & -2 \end{pmatrix}\!\begin{pmatrix}x\\ v\end{pmatrix}+\begin{pmatrix} 0 & 0\\ 1 & 1 \end{pmatrix}\begin{pmatrix}x_\ell\\ v_\ell\end{pmatrix}+\begin{pmatrix} 0 & 0\\ 1 & 1 \end{pmatrix}\begin{pmatrix}x_r\\ v_r\end{pmatrix}\right].
\label{systemeqsvdP}
\end{align}
Following the usual assumptions of phase-reduction theory \cite{kuramoto_book,pietras}, we consider that the coupling function and the differences between oscillators as given by the intrinsic frequencies are proportional to the parameter $\mu$, which measures the distance of the vdP oscillator to its Hopf bifurcation. Thus, $K = \mu \kappa$ and $\omega = \bar{\omega} + \mu \Delta \omega$, where $\mu \Delta \omega$ accounts for the deviation in the intrinsic frequency of the reference oscillator $\omega$ from the average of the distribution of intrinsinc frequencies in the lattice, $\bar{\omega}$. We thus rewrite Eq.~(\ref{systemeqsvdP}) as
\begin{widetext}
\begin{align}
\begin{pmatrix}\dot{x}\\ \dot{v}\end{pmatrix} &= \begin{pmatrix} 0 & 1\\ -\bar{\omega}^2 & 0 \end{pmatrix} \begin{pmatrix}x\\ v\end{pmatrix} + \mu \begin{pmatrix} 0 & 0\\ 0 & 1 \end{pmatrix} \begin{pmatrix}x\\ v\end{pmatrix}  + \begin{pmatrix}0\\ -x^2 v\end{pmatrix}\nonumber\\
&+ \mu \begin{pmatrix} 0 & 0\\ -2\bar{\omega} \Delta \omega & 0 \end{pmatrix} \begin{pmatrix}x\\ v\end{pmatrix} + \mu^2 \begin{pmatrix} 0 & 0\\ -(\Delta \omega)^2 & 0 \end{pmatrix} \begin{pmatrix}x\\ v\end{pmatrix}+ \mu \kappa \left[ \begin{pmatrix} 0 & 0\\ -2 & -2 \end{pmatrix}\begin{pmatrix}x\\ v\end{pmatrix} + \begin{pmatrix} 0 & 0\\ 1 & 1 \end{pmatrix}\begin{pmatrix}x_\ell\\ v_\ell\end{pmatrix} + \begin{pmatrix} 0 & 0\\ 1 & 1 \end{pmatrix}\begin{pmatrix}x_r\\ v_r\end{pmatrix}\right].
\label{systemeqsvdP2}
\end{align}
\end{widetext}

Adapting the notation in Ref.\ \cite{pietras} to our discussion, the equation above reads $\dot{\bf x} = {\bf f}({\bf x}, \mu) + \mu {\bf g}({\bf x},{\bf x}_\ell, {\bf x}_r, \mu)$, where ${\bf f}({\bf x}, \mu)$ gives the unperturbed dynamics (identical for each oscillator in the chain) in the first line on the right-hand side of Eq.~(\ref{systemeqsvdP2}), while the second line corresponds to the perturbation $\mu {\bf g}({\bf x},{\bf x}_\ell, {\bf x}_r, \mu)$, which in our case contains both the coupling to the neighboring oscillators and the randomness in the distribution of intrinsic frequencies as given by $\mu \Delta \omega$. (See e.g.\ Ch.\ 5 of Ref.~\cite{kuramoto_book} for other examples where parameter mismatches are treated as a perturbation.) The unperturbed dynamics can be written as
\begin{align}
&{\bf f}({\bf x},t; \mu)  = \bf L_0 {\bf x} + \mu \bf L_1 {\bf x} + {\bf N}_0 {\bf x}{\bf x}{\bf x},\ \label{perturbf}
 \\
&\bf L_0 = \begin{pmatrix} 0 & 1\\ -\bar{\omega}^2 & 0 \end{pmatrix}\!,\ \bf L_1 =\begin{pmatrix} 0 & 0\\ 0 & 1 \end{pmatrix}\!,\ {\bf N}_0 {\bf x}{\bf x}{\bf x} = \begin{pmatrix}0\\ -x^2 v\end{pmatrix}\!, \nonumber
\end{align}
where the nonlinear contribution corresponds to ${\bf N}_0 {\bf x}{\bf x}{\bf x} = {\bf n}_3({\bf x^{(1)}},{\bf x^{(2)}},{\bf x^{(3)}},\mu = 0)$, with
\begin{align}
&\displaystyle {\bf n}_3({\bf x^{(1)}}\!,{\bf x^{(2)}}\!,{\bf x^{(3)}}\!,\mu) = \nonumber \\
&\sum_{i_1, i_2, i_3 = 1}^2 \frac{1}{3!} \left(\frac{\partial^3 {\bf f}({\bf x},t; \mu)}{\partial x^{(1)}_{i_1} \partial x^{(2)}_{i_2} \partial x^{(3)}_{i_3}}\right)_{\!{\bf x}= {\bf 0}} x^{(1)}_{i_1} x^{(2)}_{i_2} x^{(3)}_{i_3},
\end{align}
which in our case, for $x_1 \equiv x$ and $x_2 \equiv v$, does not depend on $\mu$ anyway. The only nonzero third-order derivatives are indeed
\begin{equation}
\displaystyle \frac{\partial^3 {\bf f}({\bf x},t; \mu)}{\partial x^{(1)}_{1}\!\partial x^{(2)}_{1}\!\partial x^{(3)}_{2}}\!=\!\frac{\partial^3{\bf f}({\bf x},t; \mu)}{\partial x^{(1)}_{2}\!\partial x^{(2)}_{1}\!\partial x^{(3)}_{1}}\!=\!\frac{\partial^3{\bf f}({\bf x},t; \mu)}{\partial x^{(1)}_{1}\!\partial x^{(2)}_{2}\!\partial x^{(3)}_{1}}\!=\!-2.
\end{equation}
Quadratic terms or terms of order higher than $\mathcal{O}(|x|^3)$ do not appear in Eq.~(\ref{perturbf}), as they are absent from the  vdP oscillator dynamics [see Eq.~(\ref{systemeqsvdP2}), first line]. As for the perturbation ${\bf g}({\bf x},{\bf x}_\ell, {\bf x}_r, \mu)$, we just focus on the lowest-order terms in $\mu$ [which implies neglecting the term containing $(\Delta \omega)^2$ in the second line of Eq.~(\ref{systemeqsvdP2})]. As those terms are all linear in the state vectors of the reference oscillator and its neighbors, the expansion yields
\begin{align}
{\bf g}({\bf x},{\bf x}_\ell, {\bf x}_r, \mu) = G {\bf x} + G_\ell {\bf x}_\ell + G_r {\bf x}_r,\nonumber\\
G =  \begin{pmatrix} 0 & 0\\ -2 \bar{\omega} \Delta \omega -2 \kappa & -2 \kappa \end{pmatrix},\ \ G_\ell = G_r =\begin{pmatrix} 0 & 0\\ \kappa & \kappa \end{pmatrix}.
\end{align}

In the unperturbed dynamics at the bifurcation point, $\mu = 0$, the origin $x = \dot{x} = 0$ loses its stability. The linearized dynamics there is given by $\dot{\bf x} = {\bf L}_0 {\bf x}$, with eigenvalues $\lambda_\pm = \pm i \bar{\omega}$, right eigenvectors ${\bf u}_\pm = (1,\pm i\bar{\omega})^T$, and left eigenvectors ${\bf v}_\pm = (1/2,\mp i/2\bar{\omega})$, where the normalization has been chosen so that ${\bf v}_\pm  {\bf u}_\pm  = 1$. The solution of the linearized unperturbed system at $\mu = 0$ is thus given by
\begin{equation}
{\bf x}_0 (t) = Z e^{i\bar{\omega} t} {\bf u}_+ + Z^* e^{-i \bar{\omega} t} {\bf u}_-,
\end{equation}
where the so-called complex amplitude $Z$ is just a constant. In the standard perturbative approach, the solution of the perturbed dynamics for $0 < \mu \ll 1$ is taken to have the same form, but with a time-dependent amplitude $Z(t)$, whose typical time scale is $\tau = \mu t$. By a time-scale separation procedure due to Y. Kuramoto (see Ref.~\cite{kuramoto_book}, and also Ref.~\cite{pietras} for some other formulations giving analogous results), one can show that the complex amplitude at the lowest order of approximation is governed by the SL equation
\begin{equation}
\dot{Z}(t) = a Z(t) - b |Z(t)|^2 Z(t) + \mu [c Z(t) + c_\ell Z_\ell(t) + c_r Z_r(t)]
\end{equation}
with parameters
\begin{align}
&a = \mu {\bf v}_+ {\bf L}_1 {\bf u}_+ = \frac{\mu}{2},\\
&b = -3 {\bf v}_+ {\bf N}_0 {\bf u}_+ {\bf u}_+ {\bf u}_+^* =  \frac{1}{2},\\
&c = {\bf v}_+ G {\bf u}_+ = -\kappa + i (\Delta \omega + \kappa/\bar{\omega}),\\
&c_{\ell/r} = {\bf v}_+ G_{\ell/r} {\bf u}_+ = \frac{\kappa}{2} -i \frac{\kappa}{2\bar{\omega}}.
\end{align}
The SL equation that results is thus
\begin{align}
\dot{Z}(t) &=  \mu\left(\frac{1}{2} + i \Delta \omega \right) Z(t) - \frac{1}{2} |Z(t)|^2 Z(t)\nonumber\\
 &+K\left(\frac{1}{2}-i \frac{1}{2\bar{\omega}}\right)[Z_\ell(t) + Z_r(t) - 2 Z(t)],
\label{SLvdP}
\end{align}
where again $K = \mu\kappa$. The solution of the perturbed system is thus given by ${\bf x}(t) \approx Z(t) e^{i\bar{\omega} t} {\bf u}  + \text{c.c.}$, with $Z(t)$ satisfying Eq.~(\ref{SLvdP}), where $\text{c.c.}$ denotes complex conjugation. In terms of $z(t) \equiv Z(t) e^{i\bar{\omega} t}$, the solution to lowest order in $\mu$ becomes ${\bf x}(t) \approx z(t) {\bf u} + \text{c.c.}$, with
\begin{align}
\dot{z}(t) &= \left(\frac{\mu}{2} + i \underbrace{(\bar{\omega} + \mu \Delta \omega)}_\omega \right) z(t) - \frac{1}{2} |z(t)|^2 z(t) \nonumber\\
&+K\left(\frac{1}{2}-i \frac{1}{2\bar{\omega}}\right)[z_\ell(t) + z_r(t) - 2 z(t)].
\end{align}
In the particular case of uncoupled dynamics, $K=0$, a solution to this equation is easily obtained in polar coordinates, yielding a (stable for $\mu >0$) limit cycle of radius $\sqrt{\mu}$ with intrinsic frequency $\omega$.

The vdP oscillator can thus be approximated (close to the bifurcation point) by a SL oscillator, Eq.~(\ref{SLapp}), with $a = \mu/2+i\omega$, $b = 1/2$, and $c = 1/2-i/2\bar{\omega}$. Therefore, as discussed in the previous section, the Fourier coefficients of the coupling function $\Gamma(\psi) = A_0 + \sum_{k=1}^\infty (A_k \cos k \psi + B_k \sin k \psi)$ which appears in the phase-reduced model are
\begin{align}
A_0 &= -A_1 = \frac{c_R b_I}{b_R} - c_I = \frac{1}{2\bar{\omega}},\\
B_1 &= c_R + \frac{c_I b_I}{b_R} = 1/2,
\end{align}
all the other vanishing at this order of approximation. Thus the coupling function is
\begin{align}
\Gamma(\psi) = \frac{1}{2\bar{\omega}}(1-\cos \psi)+\frac{1}{2} \sin \psi.
\end{align}
This function is certainly not odd [$\Gamma(\psi) + \Gamma(-\psi) \neq 0$], hence the KPZ nonlinearity appears in the continuum approximation with columnar noise \cite{gutierrez}. 
This is compatible with the GSI behavior observed in the numerical results of Sec.\ \ref{secvdP}.

\section{Real-space phase correlations and morphologies}
\label{AppC}

The real-space two-point (phase-difference) correlations are defined as
\begin{equation}
G(r,t)=\langle \overline{(\phi_{j+r}(t)-\phi_{j}(t))^2}\rangle,
\label{eq:corr}
\end{equation}
with the overbar denoting spatial average across the oscillator ring and the brackets denoting the average over different disorder (i.e.\! intrinsic frequency) realizations. Due to invariance under reversal $r\to -r$, the correlations only depend on $\ell \equiv |r|$. We will be interested in cases such that $\alpha_s>1$, for which the correlation function scales as \cite{lopezphysa}
\begin{equation}
    G(\ell,t) \sim \left\{
    \begin{array}{lr}
        t^{2 \beta},& \text{if } t^{1/z} \ll \ell \ll L, \\
        \ell^{2 \alpha_\text{loc}} t^{2(\alpha-\alpha_\text{loc})/z}, & \text{if } \ell  \ll t^{1/z}  \ll L.
    \end{array}
\right.
\label{eq:corr_scal}
\end{equation}
This means that the two-point correlations keep increasing (anomalously) with time even at distances which are smaller than the correlation length, as they only saturate at $\ell^{2 \alpha_\text{loc}} L^{2(\alpha-\alpha_\text{loc})}$ when $t^{1/z} \sim L$. If $\alpha \neq \alpha_s$ with both exponents being larger than 1, as in our case, faceted anomalous scaling takes place \cite{ramasco}, and $\alpha_\text{loc} = 1$.

\begin{figure}[h!]
\includegraphics[scale=0.39]{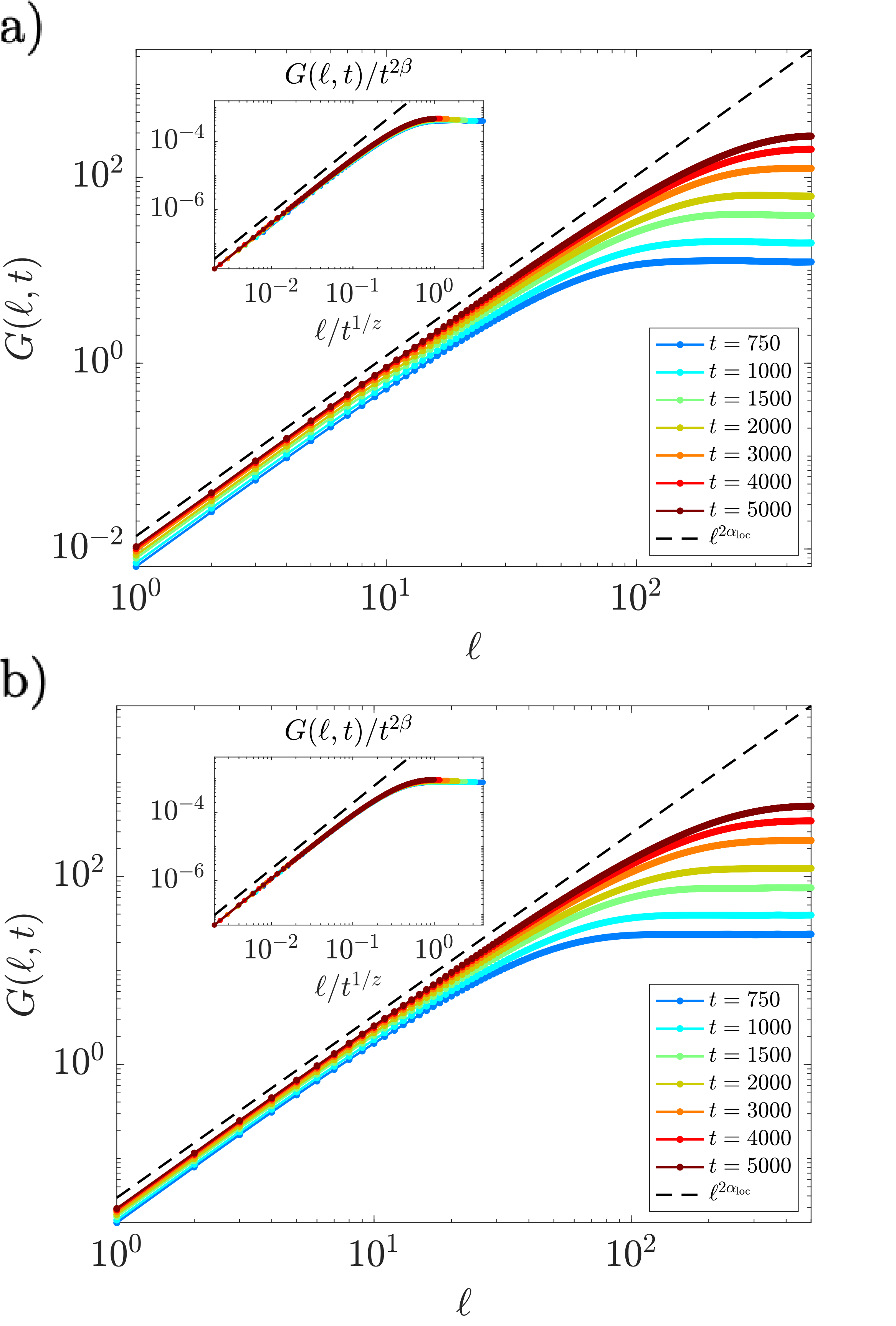}
\caption{
Two-point correlations for rings of (a) SL and (b) vdP oscillators with (inset) and without (main panel) rescaling as given by Eq.\ \eqref{eq:corr_scal}. Results shown for $\mu = 1$, $L=1000$, and $1000$ realizations. Exponent values: $\alpha = 1.07$, $z=1.37$ and $\alpha_\text{loc} = 0.97$ ($\beta = \alpha/z \approx 0.78$).}
\label{FigA1}
\end{figure}

The correlations, Eq.\ (\ref{eq:corr}), for a ring of SL oscillators are displayed in Fig.\ \ref{FigA1}(a), while analogous results are shown in panel (b) for vdP oscillators. In both cases we have chosen $\mu = 1$. While the values of the exponents $\alpha$, $\alpha_s$, and $z$ are those of Secs.\ \ref{secSL} and \ref{secvdP} (see caption), and $\beta = \alpha/z$, the value of the local roughness exponent, namely $\alpha_\text{loc} = 0.97$, has been obtained from the results here displayed, as well as similar ones obtained for other values of $\mu$ (not shown). The small discrepancy of this value with respect to the theoretically expected value may well be a finite-size effect. In the inset, the rescaled form $G(\ell,t)/t^{2 \beta}$ is shown as a function of the rescaled length $\ell/t^{1/z}$, which confirms the anomalous increase mentioned above as well as the growth $G(\ell,t) \sim \ell^{2\alpha_\text{loc}}$ for lengths smaller than the correlation length.

\begin{figure}[h!]
\includegraphics[scale=0.39]{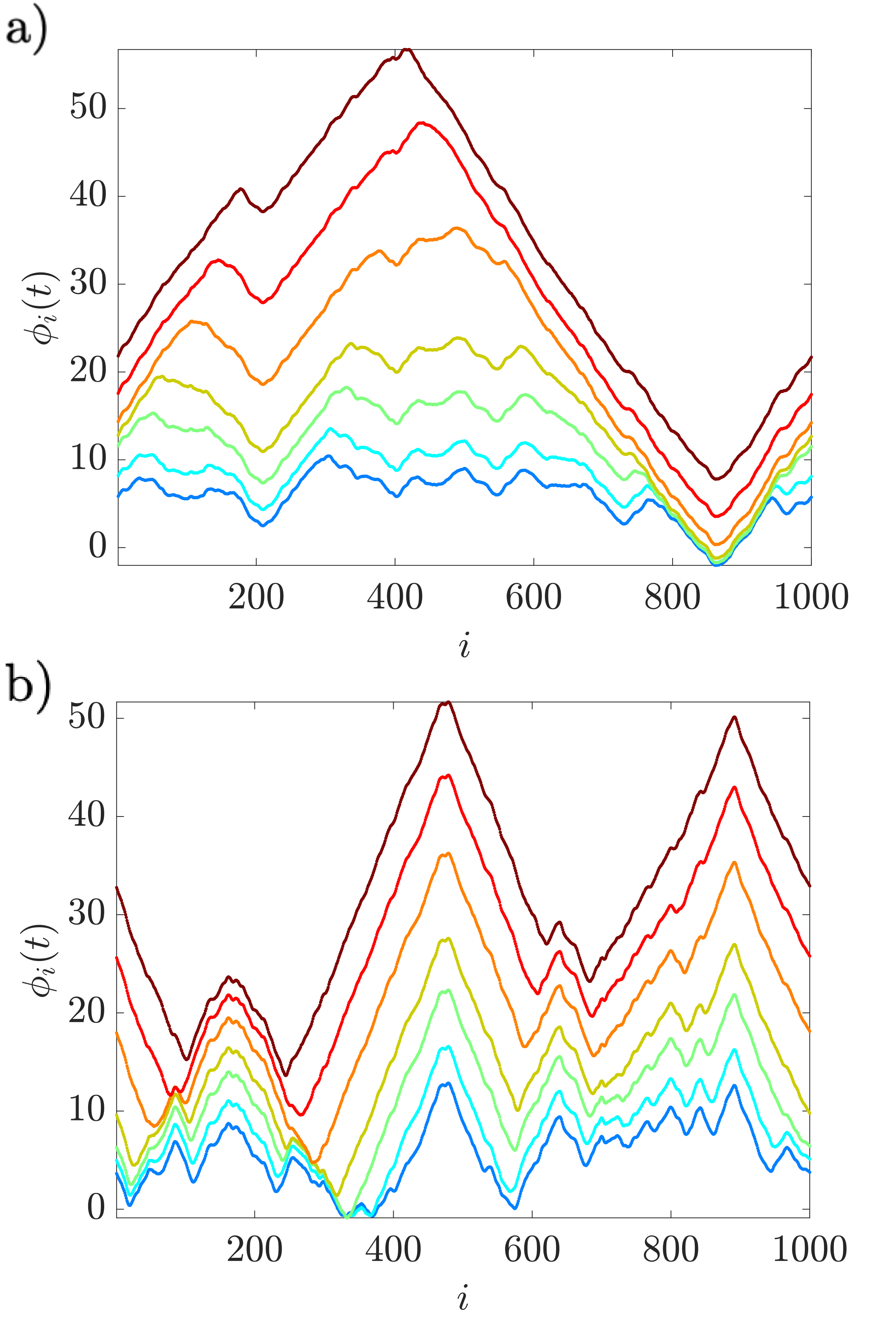}
\caption{
Representative phase profiles across time for a ring of (a) SL and (b) vdP oscillators. Results shown for $\mu = 1$ and $L=1000$. Color code as in Fig.\ \ref{FigA1}. An arbitrary vertical displacement 
has been applied for visibility purposes.}
\label{FigA2}
\end{figure}

We next illustrate the morphologies of the phase profiles in the trajectories of the rings of oscillators under consideration. Two representative trajectories are displayed in Fig.\ \ref{FigA2}, for a ring of SL oscillators in panel (a), and for a ring of vdP oscillators in panel (b), both corresponding again to $\mu=1$. Triangular facets are quite conspicuous throughout the synchronization process. They display the same characteristic coarsening dynamics previously observed in phase oscillators \cite{gutierrez,moroney} as well as in the height profile of the KPZ equation with columnar noise \cite{szendro}.

\section{Dynamical phase for rings of vdP oscillators}
\label{AppD}

As noted in Sec.\ \ref{secvdP}, even in the absence of nearest-neighbor coupling, a single vdP oscillator $(x_j(t),\dot{x}_j(t))$ behaves more and more as a relaxation oscillator for increasing values of $\mu$, moving across different regions on the limit cycle (which differs very strongly from a circle, see Fig.~\ref{fig1}, in which $\mu=1$) at different speeds \cite{strogatzbook}. Thus, the geometric phase given by $\arctan(\dot{x}_j/x_j)$ (taking into account the quadrant) no longer behaves as a valid dynamical phase variable \cite{pikovsky}. Due to the lack of an analytical expression for the isochrones (as far as we are aware), in this case we define the phase by means of a Poincar\'e section \cite{pikovsky}. Alternatively one could define the phase as $\arctan(\ddot{x}_j/\dot{x_j})$, which can be justified based on concepts derived from the differential geometry of curves \cite{osipov}, but our numerical checks do not show significant differences in the phase profiles thus obtained except at very early stages of the growth regime.

We might set the Poincar\'e section at the half line given by $x_j>0$ and $\dot{x}_j = 0$ in the phase portrait of oscillator $j$, so that, as $\dot{x}_{j}$ goes from positive to negative for positive $x_j$, the phase $\phi_j$ increases by $2\pi$ with respect to the previous crossing of the section. In point of fact, that is what we would do in the absence of coupling ($K=0$), but additional precautions need to be taken to avoid detecting spurious return times due to the `kicks' received through the coupling with neighboring oscillators for $K>0$. Specifically, two additional measures are adopted:
\begin{itemize}
\item[i)] We focus on sign changes of $\dot{x}_j$ for $x_j>\sqrt{\mu}/5$, rather than just $x_j>0$, to ensure that the oscillator goes across the `right' part of the limit cycle by the time the Poincar\'e section is crossed ---this allows for relatively strong effects due to coupling, as the limit cycle intersects the positive-$x$ semiaxis at $(x_P,0)$ for $x_P \approx 2\sqrt{\mu}$ for all $\mu$ under consideration.
\item[ii)] We only record return times for which the time elapsed since the previous one is at least one tenth of the intrinsic period $2\pi/\omega_j$, so that no rapid back-and-forth oscillation across the section may be taken for a full period.
\end{itemize}

While no set of criteria along these lines is guaranteed to work under all imaginable circumstances, we have found that the results are robust against moderate variations on our choices. Thus, a set of phase trajectories $\{\phi_j(t)\}_{j=1}^L$ are obtained from the simulations of rings of vdP oscillators, which can then be used for computing the observables described in Sec.\ \ref{secvdP}, as well as in Appendix \ref{AppC}.


\begin{thebibliography}{65}
\bibitem [{err()}]{err_exp}%
  \BibitemOpen
  \href@noop {} {}\bibinfo {note} {The errors in the exponent values can be
  considered smaller than 0.05; the exponent values must be adjusted quite
  precisely in order to satisfy the scaling constraints of $W(L,t)$, $S(k,t)$,
  and $G(r,t)$.}\BibitemShut {Stop}%
\end{thebibliography}
\end{document}